\documentclass[%
 reprint,
 amsmath,amssymb,
floatfix,
showkeys,
floatfix,
]{revtex4-2}
\usepackage{algorithm}
\usepackage{amsmath}
\usepackage[noend]{algpseudocode}
\usepackage{algpseudocode}
\usepackage{lipsum}
\usepackage{float}
\usepackage{graphicx}
\usepackage{dcolumn}
\usepackage{bm}
\usepackage{xcolor}
\usepackage{siunitx}
\usepackage{ulem}
\newcommand\hl[1]{%
  \bgroup
  \hskip0pt\color{red!80!black}%
  #1%
  \egroup
}

\newif\ifhighlight
\highlightfalse
\ifhighlight
    
\else
    \renewcommand{\hl}[1]{#1}
\fi

\begin{document}

\preprint{APS/123-QED}

\title{Design and analysis of electro-optic modulators based on high contrast gratings (HCGs) in AlGaN/GaN heterostructures}

\author{Pallabi Das}
\affiliation{Department of Electrical Engineering, IIT Bombay, Mumbai 400076, India.}
\author{Shlok Vaibhav Singh}
\thanks{P. Das and S. V. Singh contributed equally to this work. Corresponding author: S. Tallur.}
\affiliation{Department of Electrical Engineering, IIT Bombay, Mumbai 400076, India.}
\thanks{P. Das and S. V. Singh contributed equally to this work.}
\author{Siddharth Tallur}
\email{stallur@ee.iitb.ac.in}
\affiliation{Department of Electrical Engineering, IIT Bombay, Mumbai 400076, India.}


\date{\today}

\begin{abstract}
Recently High Electron Mobility Transistor (HEMT) inspired III-V electro-optic modulator topologies were proposed for realizing high speed electro-optic modulators leveraging plasma dispersion effect due to the 2D Electron Gas (2DEG) present at the III-V heterostructure interface. The 2DEG is highly confined at the interface, extending to very low depths in the bulk ($\approx$\SI{10}{nm}) and therefore has limited spatial overlap with the optical mode.
In this paper, we propose a novel modulator design to boost the 2DEG-light interaction, wherein the HEMT is embedded within a high contrast grating (HCG) mirror.
We present an analytical model extending the conventional HCG model to multi-layer structures and observe good agreement with rigorous coupled-wave analysis (RCWA). We explore the design space for identifying optimal device topology and present geometries that produce a change in reflectivity as large as \SI{70}{\%} for C- and L-band wavelengths. We also present results of sensitivity analysis and observe low variation in device performance due to geometry variation arising from device fabrication imperfections. The device platforms presented here are suitable for designing high efficiency electro-optic modulators by incorporating the HEMT HCG into a Fabry-Perot cavity.
\end{abstract}

\maketitle
\section{Introduction}
\label{intro}
\begin{figure}[bp]
\centering
\includegraphics[width=\linewidth]{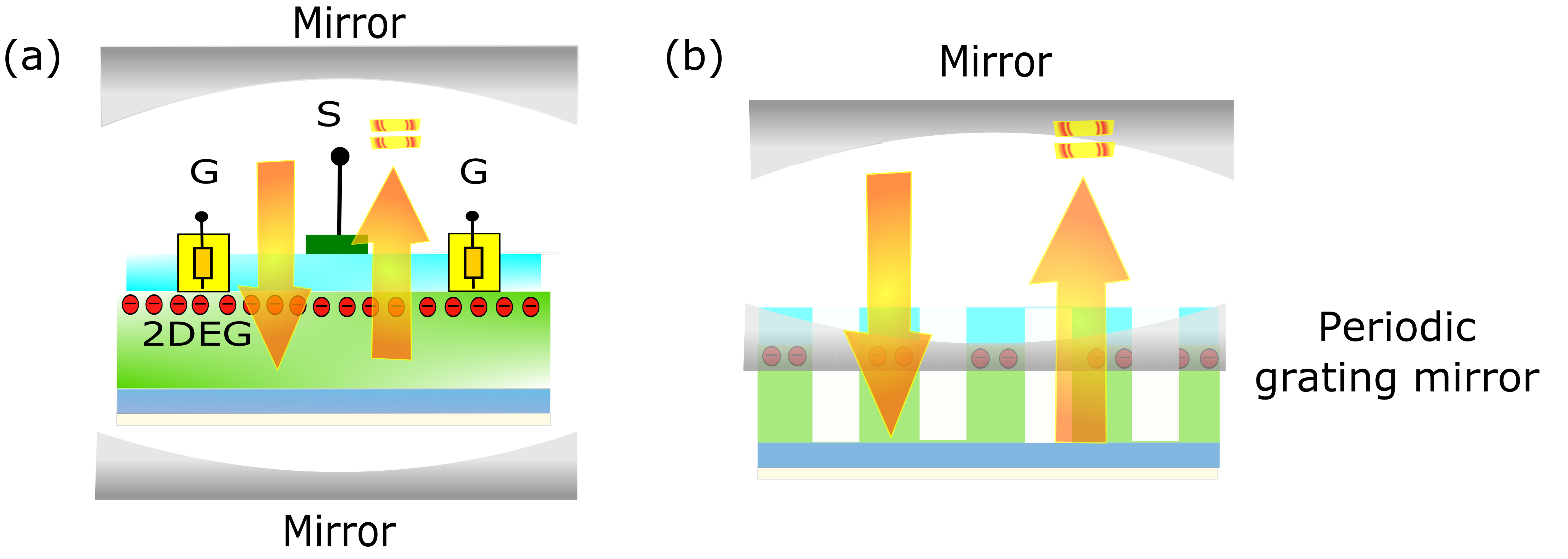}
\caption{Light interaction with 2DEG in an AlGaN/GaN modulator can be enhanced by (a) embedding the device in a Fabry-Perot cavity, or (b) incorporating the device into a high contrast grating (HCG) mirror.
The source and drain are connected to ground (G) and the modulating signal (S) is connected to the gate.
The gate voltage modulates the 2DEG concentration and thereby the reflectivity of the HCG mirror.
}
\label{fig1}
\end{figure}

Electro-optic Modulators (EOMs) find wide-spread applications beyond high speed optical communication, such as optical interconnects, generation of frequency combs, bio-sensing etc. \cite{miller2010optical,freq_comb,EO_bio1,EO_bio2}. Of various contender technologies for on-chip optical interconnects, silicon photonics has emerged as a commercially viable option owing to the electronic-photonic integration afforded by the CMOS-compatibility of silicon nanophotonic device processing \cite{lim2013review,cmos,reed_silicon}. However the speed of operation in silicon EOMs based on free carrier plasma dispersion effect in p-i-n (p-type/intrinsic/n-type layers) devices is limited by carrier recombination lifetime to $<$\SI{100}{Gbps} \cite{reed_silicon,Si_40gbps,Si_50}.
Unlike silicon, lithium niobate (LiNbO$_3$) is an electro-optically active material (Pockels effect) and chip-scale EOMs in this material platform have recently been demonstrated operating at frequencies exceeding \SI{100}{Gbps} \cite{LiNbO3,Li_cmos,wang2018nanophotonic}. However the reported modulation efficiency and bit error rates at high frequencies and the material incompatibility with CMOS processing are practical limitations that need to be overcome for this technology to become commercially viable.
Compared to the highest speed silicon CMOS transistors, gallium nitride (GaN) based High Electron Mobility Transistors (HEMTs) are suitable for high-speed and high-power applications due to the high electron mobility of the 2DEG, high saturation velocity, and large breakdown field \cite{Gan_300ghz,tang2015ultrahigh,kuzuhara2016algan}. GaN based devices not only find applications in high power devices but also in electro-optic devices owing to Pockels effect and the large band gap in GaN \SI{3.4}{eV} \cite{nitride_book,blue_gan} and several devices leveraging the Quantum-Confined Stark effect (QCSE) in multi-quantum well structures \cite{isb1,isb2} and Franz–Keldysh effect \cite{franz,franz2} in III-nitrides have been explored. Recently we reported various design topologies for EOMs leveraging the plasma dispersion effect of 2DEG in AlGaN/GaN heterostructures \cite{das2020design}, and 2DEG based EOMs operating in the UV-Vis wavelength range were reported by Wieben et al. \cite{wieben2019development}
The 2DEG is confined in an atomistically thin layer at the heterostructure interface, which results in limited optical interaction length and thereby small phase shift $\approx$ \SI{1e-5}{\radian} \cite{das2020design}. This necessitates incorporating the HEMT into a resonant cavity to boost the modulation index.  
One such manifestation is illustrated in Figure~\ref{fig1}(a), wherein the device is embedded in a Fabry-Perot cavity, an approach commonly employed in cavity opto-mechanics \cite{mim1,mim2}.

Instead of this conventional ``membrane in the middle'' approach, the modulator device could be incorporated into one of the mirrors of the Fabry-Perot cavity by designing High Contrast Gratings (HCGs), as illustrated in Figure~\ref{fig1}(b).
Conventional high-reflectivity mirrors are composed of metals such as silver or aluminum, or semiconductor Distributed Bragg Reflectors (DBRs). Membrane-type grating reflectors (MGRs) in III-V materials with reflectivity as high as $\approx$\SI{90}{\%} have been reported by Kim et al. for \SI{532}{nm} TE polarized light using AlGaN gratings \cite{kim2009algan}, and by Lee et al. for \SI{450}{nm} in GaN technology platform \cite{lee2009polarization}. 
Reports of HCG based high-reflectivity mirrors include silicon nitride mirrors reported by Stambaugh et al. \cite{mem_hcg}, and a spatial light modulator (SLM) using a surface-normal HCG resonator integrated with graphene by Sun et al. \cite{graphene_hcg}.
Sun et al. leveraged interband transitions in graphene to realize high optical modulation index with \SI{11}{\dB} extinction ratio at the resonance wavelength by applying a voltage to the monolayer of graphene atop an HCG resonator \cite{graphene_hcg}. \hl{Due to its excellent optical and electrical properties, several electro-optic devices such as photodetectors, waveguides, optical modulators have been demonstrated using graphene \cite{graphene_ref}. Spectral response of graphene-based devices range from ultraviolet to THz wavelengths. GaN being a wide bandgap material also finds applications in the visible-ultraviolet spectral range and THz devices \cite{murugapandiyan2020gan,zhang2020review,yeh2019monolithic}. Being an atomistically thin material, graphene suffers from limited spatial-interaction with light limiting modulation depth to ~\SI{2.3}{\%} \cite{cai2020graphene,graphene_hcg} similar to 2DEG EOM proposed in our earlier work \cite{das2020design}. Graphene processing is CMOS compatible, and some recent studies show adoption of GaN-on-silicon platform \cite{shi2014characteristics,dasgupta2020gan} enabling CMOS compatibility. Therefore 2DEG based EOMs are comparable with graphene modulators.}

In this paper we extend the HCG resonator concept to explore the design space for realizing 2DEG based modulators, wherein voltage-mediated modulation of the 2DEG concentration is utilized to modulate the reflectivity of the HCG resonator mirror. We develop an analytical model for multi-layer HCG structure, wherein the 2DEG layer is modeled as a thin layer sandwiched between the AlGaN and GaN films, whose refractive index is computed using Drude plasma dispersion model presented in our previous work \cite{das2020design}. We study the feasibility of realizing such modulators in AlGaN/GaN HEMTs analyzed as multi-layer HCG structures, and present geometries that produce a change in reflectivity as large as \SI{70}{\%} for C- and L-band wavelengths. We also present results of sensitivity analysis and observe low variation in device performance due to geometry variation arising from imperfections in device fabrication. The device platforms presented here are suitable for designing high efficiency electro-optic modulators by incorporating the HEMT HCG structure into a Fabry-Perot cavity.  The paper is organized as follows: section \ref{sec:overview} presents an overview of HCG based devices, and the theory of operation and analytical model for computing reflectivity. Section \ref{sec:model} describes the extended model developed for analyzing multi-layer HCGs, followed by a discussion of simulation results for design space exploration of AlGaN/GaN HEMT based HCGs in section \ref{sec:results}.

\section{Overview of HCG modulator}
\label{sec:overview}
Optical grating structures can be subdivided into three domains based on their geometric proportions in comparison to the wavelength of operation: (i) the diffraction regime, (ii) near-wavelength regime, and (iii) deep-subwavelength regime \cite{chang2012high}. The HCG is a periodic grating comprised of grating bars made of a high-index material surrounded by a low-index material, in which only zero-order diffraction is allowed. The reflectivity of HCG mirror for light incident both at normal incidence as well as at oblique incidence can be engineered by varying the grating period $(\Lambda)$, thickness ($t_g$), and duty cycle $(\eta)$ of the structure \cite{chang2012high}.
HCG relies on engineering the interference of various optical modes within the grating structure to obtain large reflectivity. Some of these modes have large spatial overlap with the boundaries of the grating bars and could be used for modulating the reflectivity by modifying the carrier concentration at the material boundaries \cite{graphene_hcg}.
While sub-wavelength gratings have been widely explored for many applications in opto-electronic devices such as vertical-cavity surface-emitting lasers (VCSELs), broadband mirrors, filters \cite{hcg_application1,hcg_mirror, hcg_filter}, electro-refraction based optical modulation incorporating multi-layer HCGs has not previously been studied.

\begin{figure}[bp]
\centering
\includegraphics[width=\linewidth]{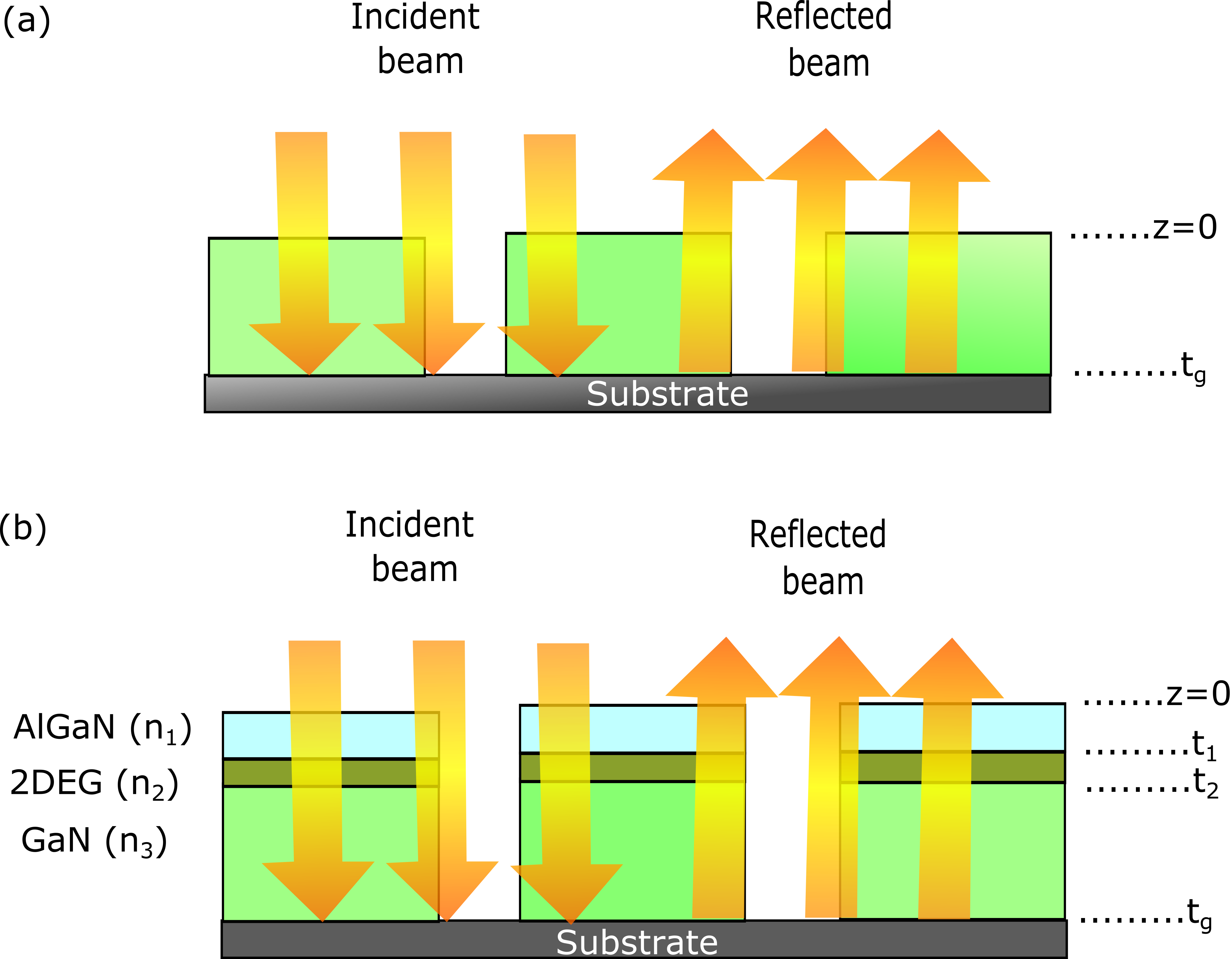}
\caption{Schematic diagram of (a) single-layer HCG, and (b) multi-layer HCG structure.}
\label{fig2}
\end{figure}


Figure \ref{fig2} shows a schematic of a single and a multi-layer HCG structure consisting of periodic array of GaN and AlGaN/GaN HEMT grating bars respectively. In order to analyze the multi-layer HCG, we first examine and analyze the single-layer HCG structure and extend the analytical model therefrom.
A plane wave when incident on the HCG bars couples to waveguide modes in the grating bars and propagates in a direction normal to the surface of the device. Each of these modes accumulate different phase as they propagate, and the interference of the waveguide modes at the input $(z=0)$ and exit planes $(z=t_g)$ is determined by the HCG thickness. A generalized analytical model of HCGs has originally been developed by Chang-Hasnain et al. \cite{chang2012high}.

\begin{figure}[!ht]
    \centering
    \includegraphics[width=\linewidth]{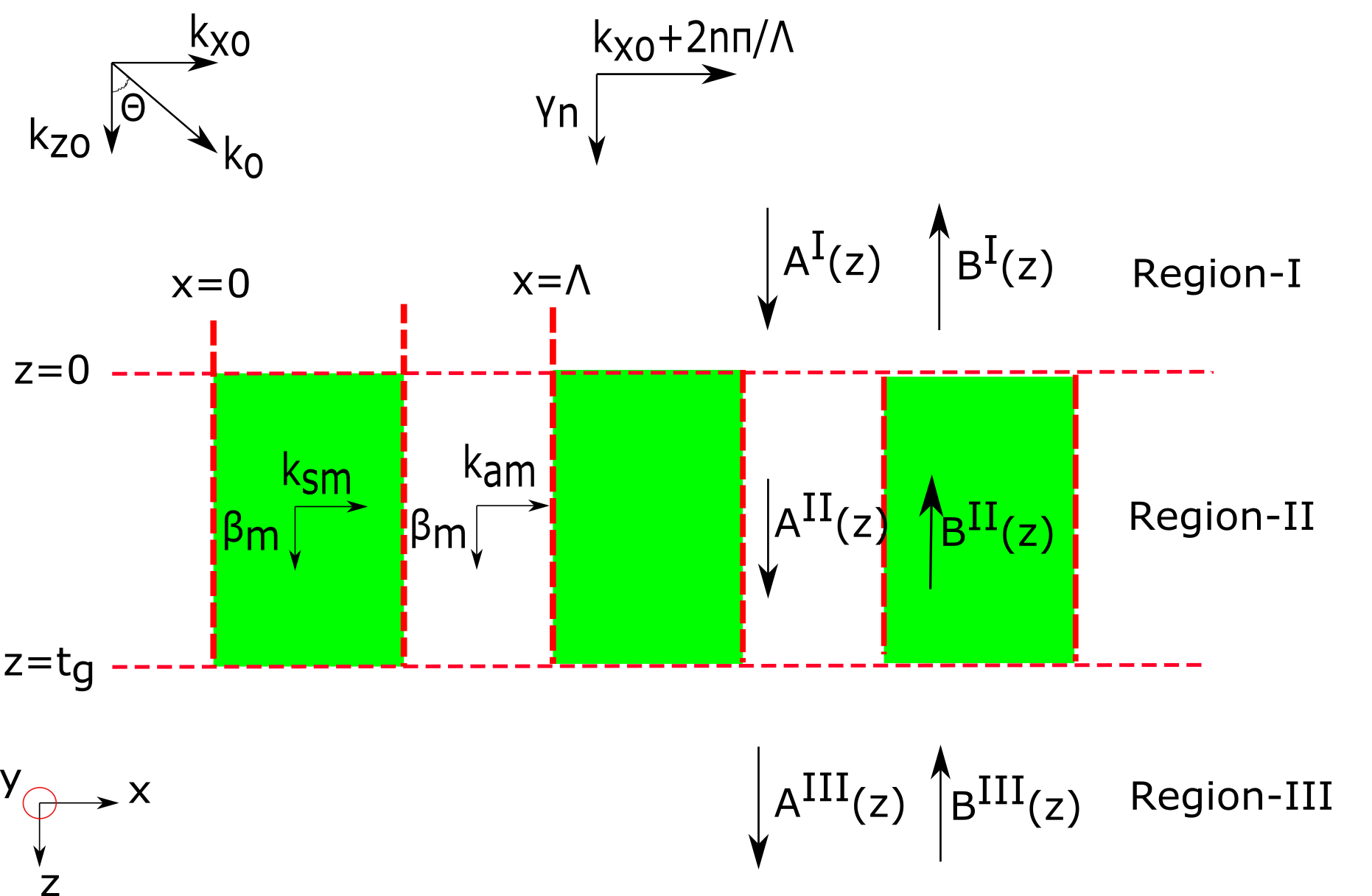}
    \caption{Illustration of single-layer HCG structure with same notations followed as \cite{chang2012high}. The structure is divided into three regions for analysis. Here $z=0$ and $z=t_g$ denote input and exit planes respectively, $A^N(z)$ and $B^N(z)$ represent forward and backward propagating coefficients respectively in the $N^{th}$ region, and $k$ and $\beta$ denote wave numbers along $X$ and $Z$ direction in corresponding region.}
    \label{fig:HCG_nomen}
\end{figure}

A schematic view of the single-layer structure is shown in Figure~\ref{fig:HCG_nomen}. We have used the same nomenclature as Chang-Hasnain et al. \cite{chang2012high} to be consistent with the symbols used. The HCG structure is divided into three regions: (i) Region I ($z<0$, cladding), (ii) Region-II ($0<z<t_g$, grating bars) and (iii) Region-III ($t_g<z$, substrate). The field profile in each region (cladding, grating, substrate) is expressed as a superposition of the eigenfunctions of the wave equations in each region. The modes are characterized by electric and magnetic field components $E^{N}_{x}$ and ${H}^{N}_{y}$ respectively, where $N$ denotes the corresponding region (I, II, or III). The electric and magnetic field in each region are expressed in terms of modal basis fields i.e $\mathcal{E}^{N}_{x}$ and $\mathcal{H}^{N}_{y}$, scaled by  combination of forward $(A^N_n(z))$ and backward $(B^N_n(z))$ propagating coefficients for the three regions \cite{chang2012high}:

\begin{multline}
H^N_y(x,z)=\sum_{n=-\infty}^{n=+\infty}[A_n^N(z)-B_n^N(z)]\mathcal{H}_{y,n}^N(x) \quad \\ E^N_x(x,z)=\sum_{n=-\infty}^{n=+\infty}[A_n^N(z)+B_n^N(z)]\mathcal{E}_{x,n}^N(x)
\label{H1}
\end{multline}

The transfer functions are represented as $T$-matrices expressed using overlap integrals $\mathbf{H}$ and $\mathbf{E}$:

\begin{gather}\label{eqn:T_I/II simplified}
\mathbf{T_{I/II}}=
\begin{pmatrix}
\mathbf{(H+E)/2} & \mathbf{(E-H)/2}\\
\mathbf{(E-H)/2} & \mathbf{(E+H)/2} 
\end{pmatrix}
\end{gather}

\begin{equation}\label{overlap}
\begin{split}
\mathbf{
H:= H_{n,m}=\dfrac{\int_0^{\Lambda}\mathcal{H}_{y,m}^{II}(x)[\mathcal{H}_{y,n}^{I}(x)]^* dx}{\int_0^{\Lambda}\mathcal{H}_{y,n}^I(x)\mathcal{H}_{y,n}^{I,*}(x)dx}
}
\\
\mathbf{
E:= E_{n,m}=\dfrac{\int_0^{\Lambda}\mathcal{E}_{x,m}^{II}(x)[\mathcal{E}_{x,n}^{I}(x)]^* dx}{\int_0^{\Lambda}\mathcal{E}_{x,n}^{I}(x)\mathcal{E}_{x,n}^{I,*}(x)dx}
}
\end{split}
\end{equation}

As the modes propagate across regions,
they either accumulate phase (in case of propagating modes) or experience amplitude extinction (for evanescent modes).
The modes at the input plane $(z=0^-)$ and output plane $(z=t_{g}^{+})$ are related by $T_{HCG}$ matrix:

\begin{gather}
\begin{pmatrix}
\mathbf{A^{I}}(0^-) \\
\mathbf{B^{I}}(0^-)  \\  
\end{pmatrix}=\mathbf{T}_{I/II}\mathbf{T}_{HCG}\mathbf{T}_{II/III}
\begin{pmatrix}
\mathbf{A^{III}}(t_g^+) \\
\mathbf{B^{III}}(t_g^+)   \\
\end{pmatrix}
\label{pmatrix}
\end{gather}

\begin{gather}
\label{THCG_psi}
\mathbf{T_{HCG}}=
\begin{pmatrix}
\psi^{-1} &0 \\
0 & \psi \\
\end{pmatrix}
\end{gather}
where,
\begin{equation}
\psi_{n,m}=
\begin{cases}
    e^{-j\beta_m t_g} & \text{$n=m$}\\
    0 & \text{$n \neq m$}
\end{cases}
\label{psi}
\end{equation}

This treatment can be extended for any number of layers by simply multiplying the $T$-matrices for all interfaces. The $n^{th}$ order reflectivity 
is obtained as $R_{n}$= $\dfrac{|\gamma_{n}|}{|\gamma_{o}|}|\mathbf{B^I}|^2$,
where $\gamma_{o}$ and $\gamma_{n}$ are the propagation constants of $0^{th}$ and $n^{th}$ order diffracted waves respectively. While the equations above are defined for infinite number of modes in either region, the behavior can be accurately captured by considering a finite number of modes \cite{chang2012high}.
The total reflectivity is obtained by adding the reflectivities of propagating modes (i.e. $\gamma_n$ is real-valued):

\begin{equation}
    R_{total} = \sum_{n : \gamma_n \ni \mathbb{R}}^{} R_n = \sum_{n : \gamma_n \ni \mathbb{R}}^{} \dfrac{|\gamma_{n}|}{|\gamma_{o}|}|\mathbf{B^I}|^2
\end{equation}



\section{Modeling and simulation setup}
\label{sec:model}
\subsection{Analytical model for multi-layer HCG}

\begin{figure}
\centering
\includegraphics[width=\linewidth]{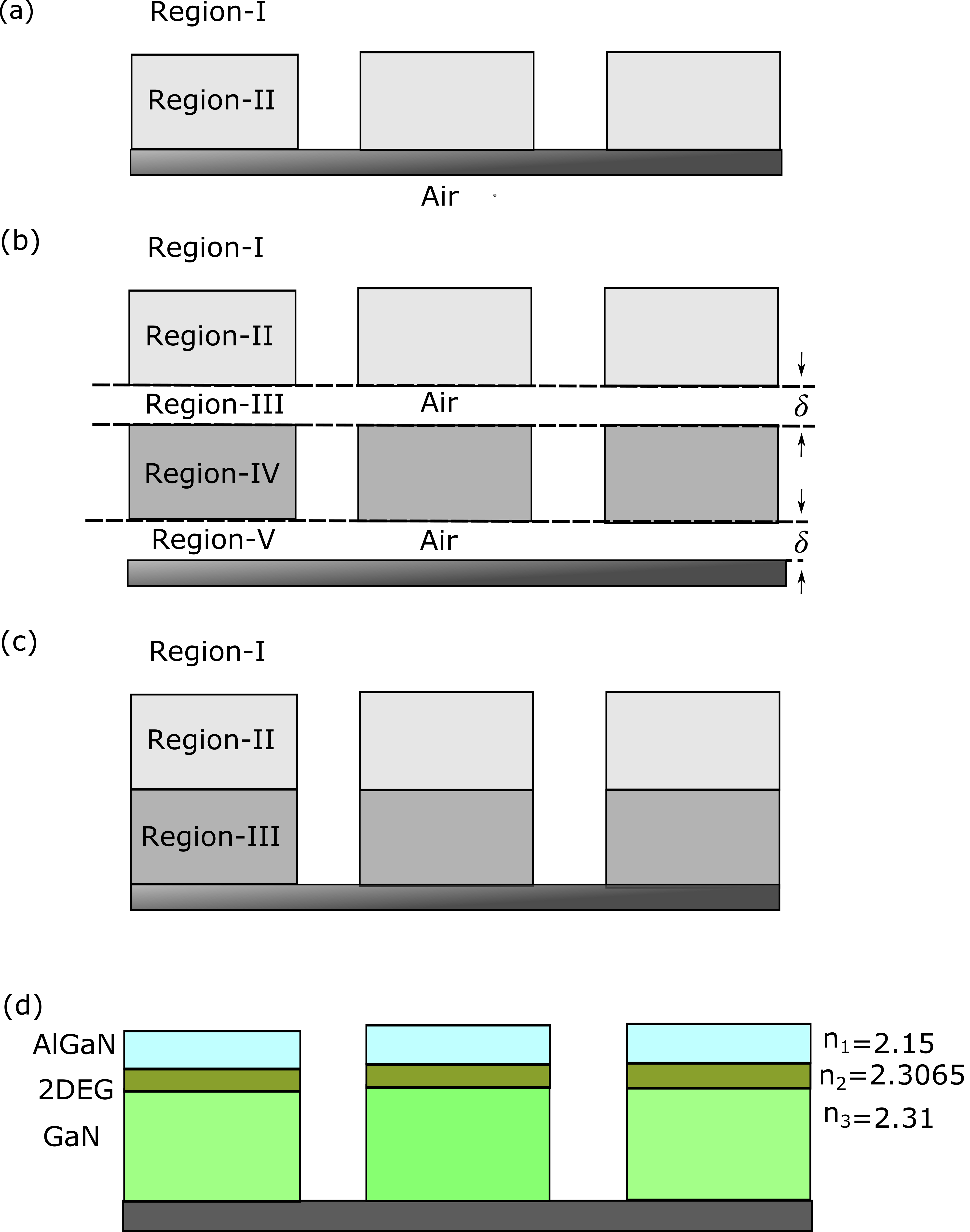}
\caption{(a) Single-layer HCG. (b) We represent a multi-layer HCG as two stacked single-layer HCGs separated by air region of thickness $\delta$. (c) In the limiting case $\delta \to 0$, the air slab diminishes and the structure resembles a multi-layer HCG. (d) Schematic representation of 3-layer AlGaN/GaN HCG, with 2DEG modeled as a separate layer (not drawn to scale).}
\label{single_multi}
\end{figure}

To analyze a multi-layer structure we start with the model for a single-layer HCG shown in Figure \ref{single_multi}(a). We consider two HCG structures with air-slabs of thickness $\delta$ inserted between the two HCG structures and the bottom substrate as shown in Figure \ref{single_multi}(b). These air slabs are denoted as Region III and Region V.
For air-substrate interface we use Fresnel equation and for air-HCG interfaces, we formulate $T-$matrices following the methodology described in previous section. After formulating the $T-$matrices for all interfaces, we solve for the limiting case $\lim {\delta \to 0}$, such that the air slab diminishes and the structure corresponds to a multi-layer HCG structures as shown in Figure \ref{single_multi}(c).
For grating bars with high aspect ratio (thickness : width) exceeding $1:1$, the entries in the propagation matrix ($\psi$, details in equation~\eqref{psi}) corresponding to evanescent modes grow exponentially large or become infinitesimally small, causing computation errors and numerical instability in calculating the inverse of the propagation matrix required to calculate the corresponding $T-$matrix (see equation~\eqref{THCG_psi}). To prevent this, we truncate the values in the propagation matrices such that the largest values are cut-off at \SI{1e3}{} and the smallest values are cut-off at \SI{1e-3}{}. These optimized threshold values are chosen based on benchmarking against Rigorous Coupled Wave Analysis (RCWA). 
We observe that this approach has an error $<$\SI{0.01}{\%} as compared to RCWA \cite{manual} if a large number of modes ($>$\SI{100}{}) are considered in the HCG analysis. While HCG structures can be designed for both TE and TM polarized light, we limit our analysis to TM polarized light and normal incidence in the discussion below. The same methodology is also applicable to TE polarized light as well as arbitrary angles of incidence.

For the AlGaN/GaN based HCG device, we have modeled a multi-layer HCG wherein the 2DEG layer is modeled as a thin slab sandwiched between the AlGaN and GaN films. An illustration of the AlGaN/2DEG/GaN HCG structure is shown in Figure \ref{single_multi}(d).
The AlGaN layer is the top layer in the HCG bar, with thickness \SI{25}{nm} and refractive index of \SI{2.15}{} at \SI{1.55}{\micro m} wavelength. In our previous work \cite{das2020design} we have analyzed electro-optic modulation due to plasma dispersion effect of 2DEG in an AlGaN/GaN HEMT, wherein we observed that the electron wavefunctions in the 2DEG are highly localized within a distance of \SI{10}{nm} at the AlGaN-GaN interface, which is also consistent with other reported calculations \cite{qin2018modeling}. Therefore the thickness of the 2DEG layer in the multi-layer HCG is considered as \SI{10}{nm}. We have also calculated electro-refraction induced change in refractive index of magnitude \SI{3.5e-3}{} for change in gate voltage $\Delta V_g\approx$ \SI{6}{V} \cite{das2020design}. Therefore the refractive index of the 2DEG layer is set to be the same value as GaN (\SI{2.31}{}) in OFF state (i.e when the 2DEG is not present) and \SI{2.3065}{} in ON state (i.e when 2DEG is present).

\subsection{Role of refractive index of substrate}
\label{ssec: substrate effect}
\begin{figure}[!ht]
\centering
\includegraphics[width=\linewidth]{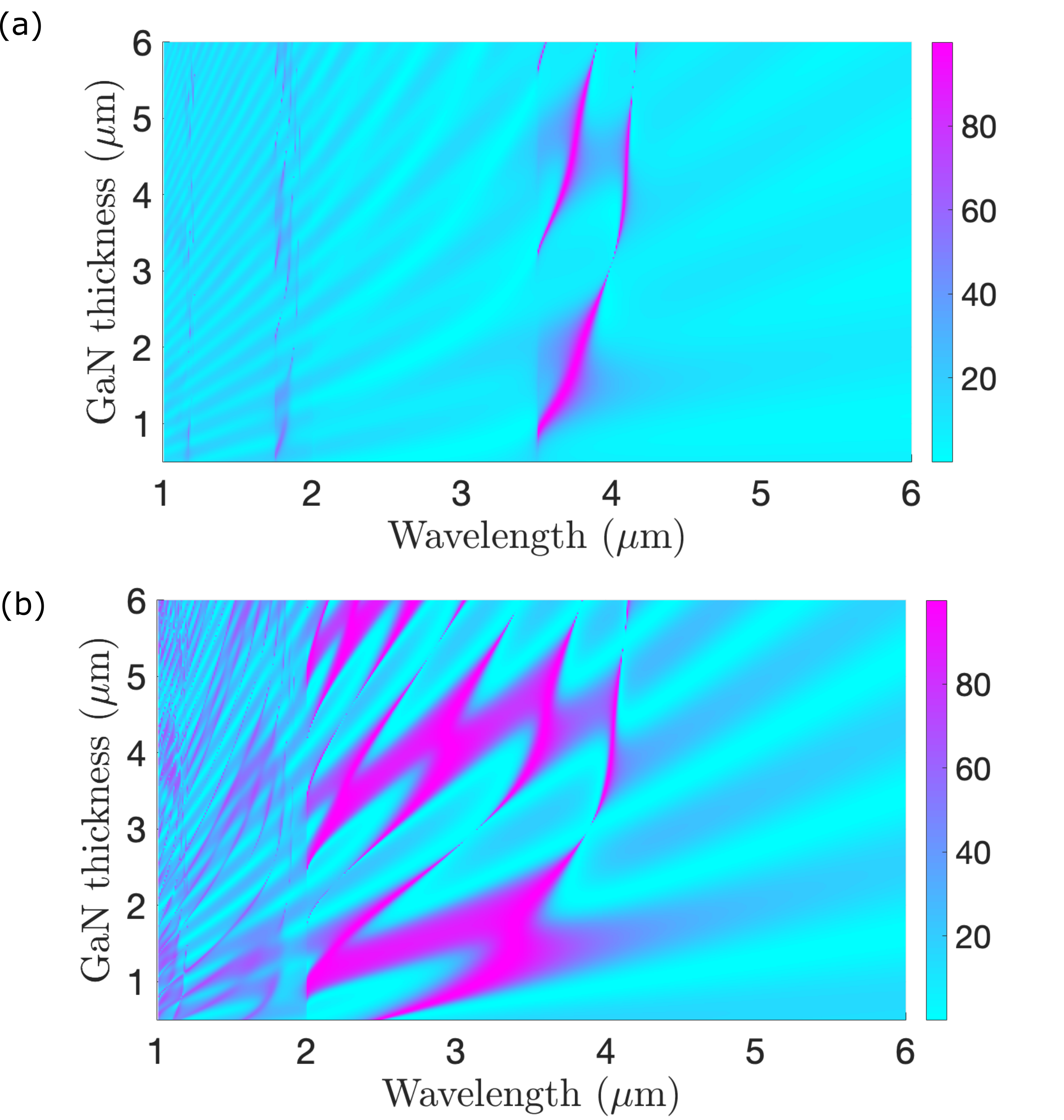}
\caption{HCG reflectivity shown as a contour plot for various values of wavelength and GaN thickness for (a) sapphire, and (b) air substrates. The color denotes the reflectivity expressed as percentage. The suspended gratings (i.e. air substrate) show large number of resonances in the telecom wavelength range and are more suitable for exploring modulator designs.}
\label{substrate_ref}
\end{figure}

We consider two material options for the substrate: (i) sapphire (conventionally used in AlGaN/GaN HEMTs), and (ii) air (suspended grating structures \cite{zhou2018high}). The performance (reflectivity) of HCG gratings in ON state for various values of GaN thickness and wavelength of light for both substrates are shown in Figures~\ref{substrate_ref}(a) and (b) respectively.
The optimal values of grating period, duty cycle and air gap used in these simulations are provided in Table~\ref{tab:my_label}. We infer from Figure~\ref{substrate_ref} that the suspended HCG structure has large number of resonances with high reflectivity (especially in the telecom wavelength range), and is better suited for low-loss modulation of TM polarized free-space light beam. We consider suspended gratings in subsequent discussion in this paper.

\begin{table}[!bp]
\caption{Design parameters: Optimal values and range explored for sensitivity analysis} \label{tab:my_label}
\setlength\tabcolsep{0pt} 
\footnotesize\centering

\smallskip 
\begin{tabular*}{\columnwidth}{@{\extracolsep{\fill}}cccc}
\toprule
  Parameter &  Optimal value & Lower limit & Upper Limit \\
  \hline \hline
 GaN thickness  & \SI{3.1}{\micro m} & \SI{2.8}{\micro m} & \SI{3.4}{\micro m} \\\hline
       Duty cycle  & \SI{0.695}{} & \SI{0.665}{} & \SI{0.725}{}\\ \hline
       Grating period  & \SI{3}{\micro m} & \SI{2.9}{\micro m} & \SI{3.1}{\micro m}\\ \hline
\end{tabular*}
\end{table}

\section{Results and Discussion}
\label{sec:results}

\begin{figure}[!ht]
     \centering
     \includegraphics[width=\linewidth]{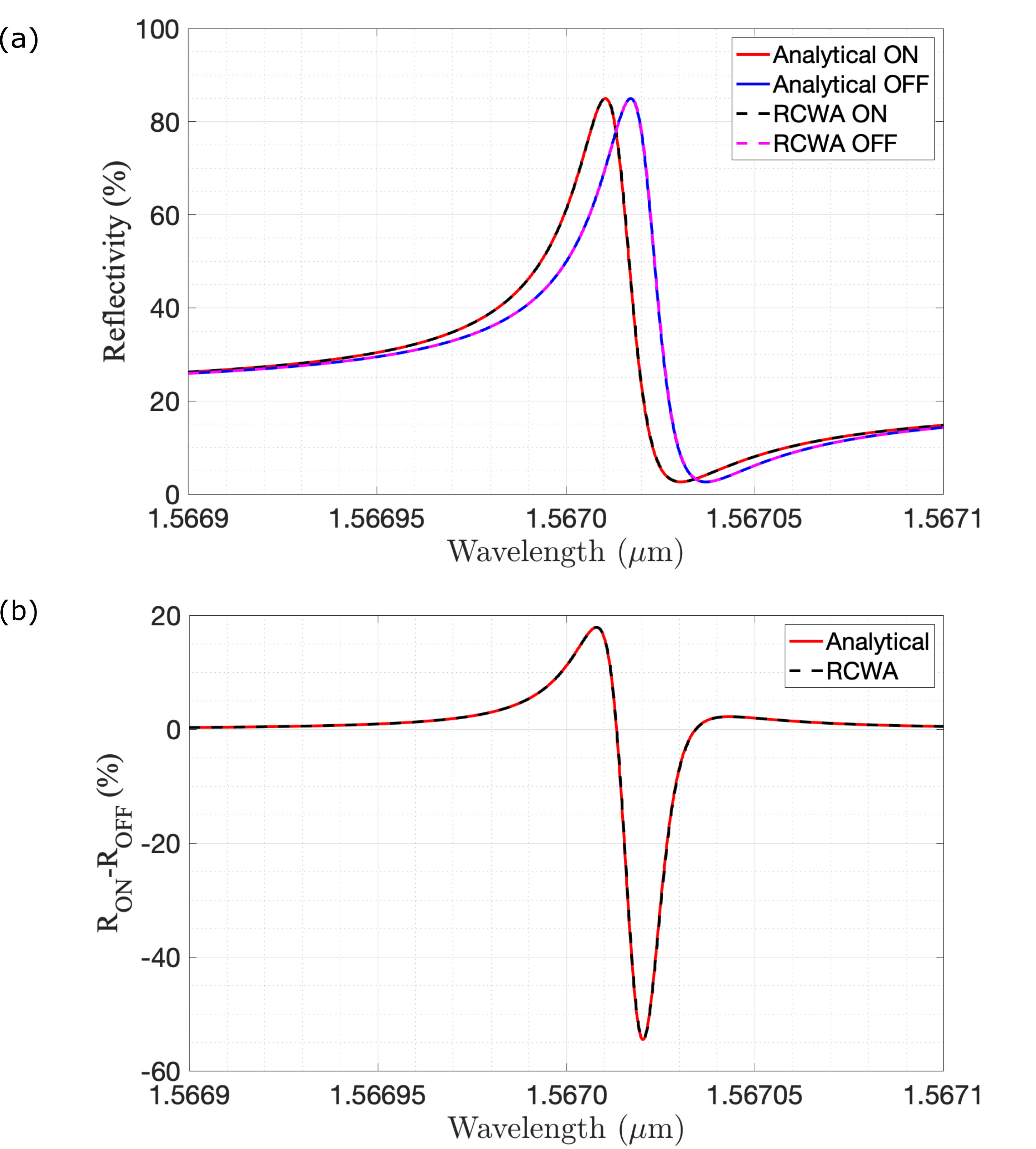}
	\caption{(a) Variation of reflectivity for AlGaN/2DEG/GaN HCG structure showing excellent agreement of the analytical model with RCWA. The resonance wavelength shows a red-shift when the device is switched from ON state to OFF state. Results are shown for optimal geometry parameters presented in Table \ref{tab:my_label}. (b) \hl{Shows modulation depth $(R_{ON}-R_{OFF})$ (i.e. difference between reflectivity in “ON” state and “OFF” states) of \SI{17.35}{\%} at an operating wavelength $=$\SI{1567.01}{nm}, whereas when the input light wavelength $=$\SI{1567.017}{nm}, the modulation depth is \SI{53.51}{\%}. In order to obtain maximum modulation depth, we should operate the modulator at \SI{1567.017}{nm} wavelength, which is the resonance wavelength in the “OFF” state.}}
	  \label{showcase_resonance}
\end{figure}

The optimal set of parameters for obtaining high reflectivity HCG mirrors for \SI{1.55}{\micro m} wavelength TM polarized light are shown in Table \ref{tab:my_label}. Graphs showing variation of reflectivity with wavelength for the optimal design parameters in ON and OFF states of the modulator are shown in Figure~\ref{showcase_resonance}(a).
We observe a red-shift in the resonance wavelength when the device is switched from ON to OFF state due to variation in the refractive index of the 2DEG. Although the thickness of the 2DEG layer is only \SI{10}{nm} and the change in refractive index of this layer is only \SI{3.5e-3}{}, we observe a large shift in reflectivity, as high as \SI{17.35}{\%}. \hl{This shift in resonance wavelength is due to variation in the refractive index of the 2DEG. Considering an operating wavelength $=$\SI{1567.01}{nm}, Figure~\ref{showcase_resonance}(b) shows modulation depth $(R_{ON}-R_{OFF})$ (i.e. difference between reflectivit in “ON” state and “OFF” states) of \SI{17.35}{\%} whereas when the input light wavelength = \SI{1567.017}{nm}, the modulation depth is \SI{53.51}{\%}. In order to obtain maximum modulation depth, we should operate the modulator at \SI{1567.017}{nm} wavelength, which is the resonance wavelength in the “OFF” state}. 

 \begin{figure}[!ht]
	\includegraphics[width=\linewidth]{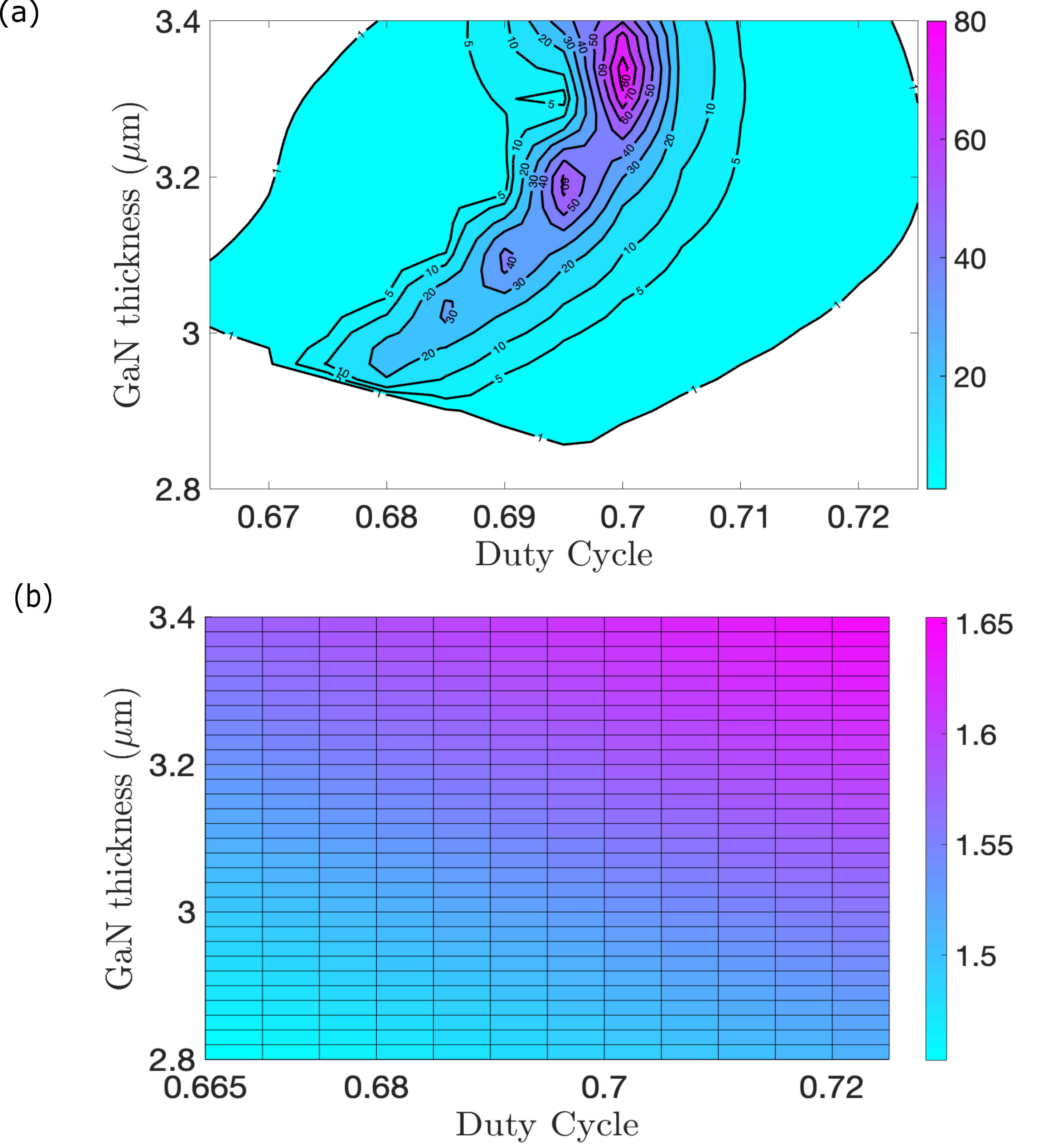}
		\caption{Results of sensitivity analysis showing variation in (a) peak reflectivity at ON-state resonance, and (b) ON-state resonance wavelength as contour plots. The grating period is fixed at \SI{3}{\micro m}.}
		\label{contour_peak_example}
\end{figure}

To analyze the sensitivity of the peak reflectivity and resonance wavelength to variations in geometric parameters arising from fabrication imperfections, we vary the GaN layer thickness, grating period and duty cycle within a range shown by the lower and upper limits for each parameter in Table~\ref{tab:my_label}. The ranges are chosen based on typical process variation windows. The AlGaN thickness is not varied in our sensitivity analysis, as the AlGaN layer is typically grown using molecular beam epitaxy (MBE) process that has excellent parameter control.
For this analysis we vary the duty cycle 
ranging from \SI{0.665}{} to \SI{0.725}, keeping the grating period fixed at \SI{3}{\micro m}.
The GaN layer thickness is varied from \SI{2.8}{\micro m} to \SI{3.4}{\micro m}.
The peak ON-state reflectivity and corresponding resonance wavelength are shown as contour plots in Figures~\ref{contour_peak_example}(a) and (b) respectively.
We observe that significant reflectivity ($>$\SI{50}{\%}) is obtained for the HCG design in a small portion of the design space. The white-space in Figure~\ref{contour_peak_example}(a) corresponds to the case where change in reflectivity ($|R_{ON}-R_{OFF}|$) is $<$\SI{1}{\%}.
The resonance wavelength $(\lambda_0)$ is observed to vary linearly with the duty cycle $(\eta)$ and GaN thickness $(t_{GaN})$, as seen in Figure~\ref{contour_peak_example}(b). We perform linear regression using MATLAB curve-fitting toolbox to obtain the following equation with $R^2 = 0.9986$:
\begin{equation}
\lambda_0 [\mu m] = 0.05784+0.2141 \times t_{GaN} [\mu m] + 1.196 \times \eta
\end{equation}

\begin{figure*}[ht]
	\includegraphics[width=0.9\linewidth]{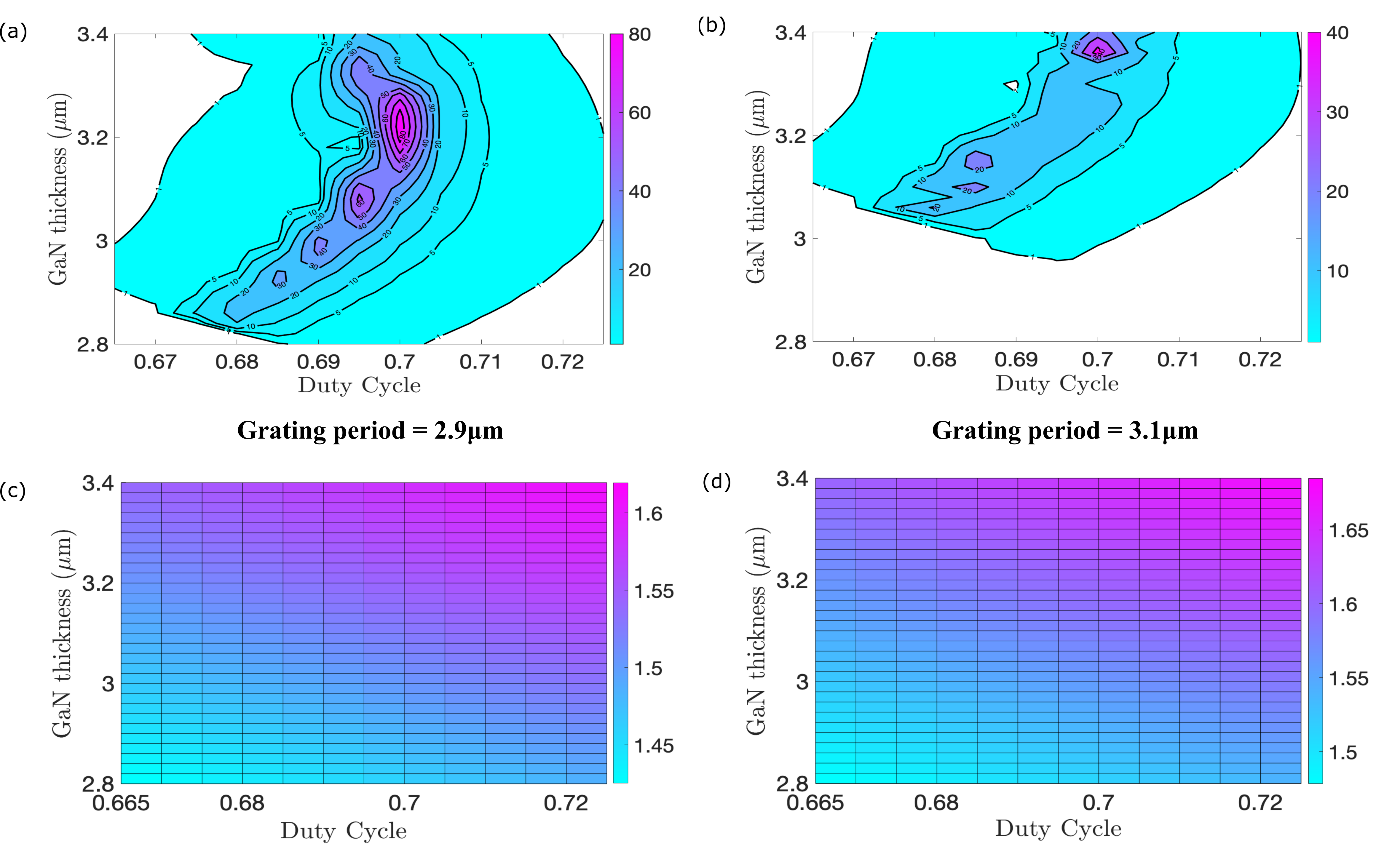}
		\caption{Results of sensitivity analysis showing variation in peak reflectivity at ON-state resonance for grating period (a) \SI{2.9}{\micro m} and (b) \SI{3.1}{\micro m}. The corresponding contour plots for variation of resonance wavelength are shown in panels (c) and (d) respectively.}
		\label{contour_peak}
\end{figure*}

Next we perform the same study for two other values of grating period $(\Lambda)=$ \SI{2.9}{\micro m} and \SI{3.1}{\micro m}. The corresponding results are shown in Figure~\ref{contour_peak}.
We observe that while a significant portion of the design space yields reflectivity $>$\SI{20}{\%}, this comes at the cost of the resonance wavelength exiting the C-band and entering S- and L-bands. The resonance is observed to be more sensitive to the grating period and duty cycle than GaN thickness, suggesting that tight control of lithography parameters (e.g. achievable with electron-beam lithography) is essential to realize high efficiency modulators based on the architecture proposed here.

\section{Conclusion and future work}
In summary, we have explored the design space for electro-optic modulators based on AlGaN/GaN HEMTs designed as multi-layer HCG structures. We present an extension of the conventional single-layer HCG model to analyze the performance of these structures. The analytical model thus developed shows excellent agreement with RCWA.
Even though the 2DEG layer is highly localized at the AlGaN-GaN interface, and has calculated spatial extent of merely \SI{10}{nm} with small change in refractive index (\SI{3.5e-3}{}) due to applied gate voltage, incorporating the HEMT into an HCG structure results in large change in reflectivity ($>$\SI{50}{\%}) by switching the 2DEG from ON (saturation) to OFF (depletion) state. This boost in modulation efficiency is significant improvement from HEMT based modulator architectures presented in our previous work \cite{das2020design}.
The HCG resonance wavelength showing a linear dependence on the GaN thickness and duty cycle ($\eta$) of the gratings. As seen in Figure~\ref{contour_peak}, the proposed structure is very sensitive to variation in critical dimensions for the gratings with a large portion of high reflectivity region in the design space present in a narrow range within $\eta =$ \SI{0.69}{} and \SI{0.705}{}, allowing  a variation of only \SI{20}{nm} in critical dimension. The reflectivity variation with change in GaN thickness is more acceptable with reflectivity consistently $>$\SI{20}{\%} observed for a variation of larger than \SI{400}{nm} in GaN thickness. 
Another aspect to be considered in determining the feasibility of fabricating such structures is the necessity of suspended gratings, as discussed in section~\ref{ssec: substrate effect}.
AlGaN/GaN heterostructures grown by Metal Organic Vapor Phase Epitaxy (MOVPE) can be etched by Photo-Electro-Chemical (PEC) etching in various dilute electrolytes \cite{etch1,etch2}. However other non-idealities associated with etching processes such as etch-loading for high aspect ratio structures, sidewall angle and roughness etc. need to be investigated. Our future work will focus on fabrication process development and optimization for realizing the suspended AlGaN/GaN HCG structures presented here and experimentally validating the modulator performance based on plasma dispersion effect of 2DEG. 

\begin{acknowledgments}
The authors thank Prof. Swaroop Ganguly and Prof. Dipankar Saha at IIT Bombay for their valuable inputs on TCAD simulations, and Ms. Kiran Dhope at IIT Bombay for help with conducting preliminary simulations. The authors also thank Microelectronics Computation Lab (MCL) at IIT Bombay for providing access to Silvaco ATLAS. 
\end{acknowledgments}


\bibliography{ref}   
\end{document}